\documentclass[useAMS,usenatbib]{mn2e}
\usepackage {graphicx}
\usepackage {times}

\title{A Disc-Corona Model for
    a Rotating Black Hole}
\author[Xiao-Long Gong,Li-Xin Li and Ren-Yi Ma]{Xiao-Long Gong$^{1}$\thanks{E-mail:
gongxiaolong@mail.bnu.edu.cn},Li-Xin Li$^{2}$ and Ren-Yi Ma$^{3}$ \\
$^{1}$ Department of astronomy, Beijing Normal University, Beijing,
100875, P. R. China\\
$^{2}$ Kavli Institute for Astronomy and Astrophysics, Peking University ,Beijing,
100875, P. R. China\\
$^{3}$ School of Physic and Mechanical \& Electrical Engineering, Xiamen University, Xiamen,
361005, P. R. China}
\begin{document}

\date{}

\pagerange{\pageref{firstpage}--\pageref{lastpage}} \pubyear{2011}
\maketitle

\label{firstpage}

\begin{abstract}

\noindent
We propose a disc-corona model in which a geometrically thin, optically
thick disc surrounds a Kerr black hole, and magnetic fields exert
a time-steady torque on the inner edge of the accretion disc. The analytical
expression of the total gravitational power is derived from the thin-disc dynamics equations by using this new boundary condition.
It is shown that the magnetic torque can considerably enhance the amount of energy released in
the disc-corona system. Furthermore, the global solutions of this disc-corona system are
obtained numerically. We find that the fraction of the power dissipated into the corona in the total for such disc-corona system increases with the increasing dimensionless black
hole spin parameter $a_\ast $, but is insensitive on the $\Delta \varepsilon
$ which is the additional radiative efficiency parameter relevant to
magnetic torque, for $\Delta\varepsilon > 1$. In addition, the emerged spectra from this disc-corona
system are simulated by using Monte-Carlo method, and the effect of the different parameters on the output spectra is discussed.

\end{abstract}


\section{Introduction}
\noindent
It is believed that the black hole (BH) accretion disc is an effective model for explaining the high energy radiation in astrophysics. The
standard accretion disc (SSD) model was proposed by \citet{b30}, in which the disc is geometrically thin and optically thick. The SSD is
widely used in modeling the spectral energy distribution (SED) of AGN. The
general relativistic SSD model has been investigated in detail by Novikov
{\&} Throne (1973, hereafter NT73), and Page {\&} Throne (1974, hereafter PT74). In this SSD model, it has
been assumed that there is no stress at disc's inner edge, i.e. the
so-called ``no-torque inner boundary condition''.

However, the ``no-torque inner boundary condition'' was questioned by \citet{b10}
 on the basis that magnetic fields are the likely agent of the torque
in the discs \citep{b2}. \citet{b1} proposed that
magnetic fields connecting the disc to the plunging region can exert stresses on the
inner edge of an accretion disc around a black hole, and they recomputed the
relativistic corrections to the thin-disc dynamics equations when these
stresses take the form of a time-steady torque on the inner edge of the
disc. \citet{b6} also noted that, within the confines of a
highly-idealized model of inflow dynamics, this torque can considerably
enhance the amount of energy released in the disc.

Apart from SSD, advection dominated accretion flow (ADAF) is another
important model of the accretion flow \citep{b23,b24}. It is believed that black hole X-ray binaries
present various spectral states, most notably the low/hard state, high/soft
state and intermediate state. The SSD has been very successful in describing
the high /soft states of Galactic black hole candidates (GBHCs), but their
low/hard state characterized by power-law--type spectra was problematic to
the SSD. The ADAF model is quite successful in reproducing the hard spectra
of GBHCs, as well as those of low-luminosity AGNs (LLAGNs),
however it also has a several problems. Thus we cannot expect to interpret
the various components of spectra based on one accretion mode.

The power-law spectra of GBHCs is generally explained by Comptonization
of softer photons by hot electrons from a magnetic corona on an accretion
disc. \citet{b15} proposed a high-temperature corona model
analogous to solar corona, and the effects on the disc when part of the
dissipation occurs in the corona were first discussed by \citet{b9}
. In the accretion disc-corona scenario, an optically thin hot
thermal corona is located above the surface of disc. A fraction of soft
photons, which are from the cold disc, are Compton up-scattered to X-ray photon
by hot electrons. In this model the corona can explain the power-law X-ray
spectra very well, and reprocessing of the coronal X-rays by the cold disc gives
rise to the observed emission lines naturally. The iron $K_{\alpha}$
fluorescence line provides us a diagnostic of the geometry of the accretion
flow and the property of the space-time around the BH. Recent work on the disc-corona model can be found in, e.g. \citet{b21,b16,b5}.

\citet{b30} used the famous ``$\alpha
$-prescription'' to deal with viscosity, but the
physical process leading to viscosity and turbulence in the disc remain unclear. The magnetic rotation instability (MRI) of weak magnetic fields in accretion discs is thought to play an important role in the evolution and dynamics of
astrophysical accretion discs. \citet{b2} argued that this
instability should have a rapid growth rate of the order of the orbital
frequency $\Omega $, resulting in a greatly-enhanced effective viscosity
that is able to transport angular momentum outward. On the other hand, the
magnetic fields generated in the SSD are strongly buoyant, and a fraction of
the magnetic energy is transported vertically to heat the corona above the disc.

The multi-wavelength observations of nearby LLAGNs have revealed that the
SEDs of LLAGNs and of GBHCs in the low/hard state possess many similarities,
e.g. flat, compact radio cores with high brightness temperatures, and a hard X-ray
power-law with high a energy cut-off. \citet{b21} pointed out
that there should be a common accretion mode for these low luminosity black
holes, and proposed a new model for low-luminosity black holes, in which a
SSD at low accretion rates dissipates a large fraction of its gravitational
energy in a magnetic corona.

Motivated by the these results, in this paper we shall investigate a disc-corona
model, in which a geometrically thin, optically thick disc surrounds a
rotating Kerr BH, a magnetic field connecting the plunging region and the disc exerts a non-zero torque at the inner boundary of the disc, and the corona is assumed to be heated by the reconnection
of the magnetic fields generated by buoyancy instability in the disc. This
paper is organized as follows. We describe the model in section 2, where in a
magnetic torque is exerted on the inner edge of the accretion disc. In section
3 the global solutions of this disc-corona system are obtained, and the
effects of different parameters on the fraction of accretion power
dissipated into the corona are discussed. In section 4 we simulate the emerged
spectra by using a Monte Carlo method. Finally, in section 5, is the brief
discussion. Throughout this paper the geometric units $ G = C = K = 1$ are used.

\section{Description of Model}
\noindent
In our accretion disc-corona system, a geometrically thin and optically
thick disc is sandwiched by a magnetic corona, and part of
the gravitational energy of the accreted matter is released in the hot corona. The
general relativistic model for a steady, axisymmetric, and thin Keplerian
disc around a Kerr black hole has been described in detail by NT73 and PT74 . In their model, it has been
assumed that there is no stress at disc's inner edge. The equation of
angular momentum conservation is given as follows

\begin{equation}
\label{eq1}
[\dot {M} L^ + - 2\pi e^{\nu + \psi + \mu }W]_{,r} = 4\pi e^{\nu + \psi +
\mu }Q L^ + ,
\end{equation}

\noindent
where $\dot {M} $ is the accretion rate of the disc,$\upsilon$,
$\psi$, $\mu$ are the metric coefficients, $W$ is
integrated shear stress, $L^{+}$ is the specific angular momentum of a particle in the disc.

The total gravitational power dissipated in unit surface area of the
disc-corona system $ Q $ is given by

\begin{equation}
\label{eq2}
Q = (\dot{M}/4\pi) e^{-(\upsilon + \psi + \mu)}f,
\end{equation}

\noindent
where the function of radius $f$ is defined by PT74. In the case of ``no-torque
inner boundary condition'', the boundary condition on $f$at the radius $r_{ms}$ of marginally stable orbit
is $f_{ms} = 0$.

If it is taken into account that magnetic stresses exert a torque on the
inner edge of the accretion disc, the appropriate boundary condition at
$r_{ms}$ is expressed as \citep{b1}

\begin{equation}
\label{eq3}
f_{ms} = \frac{3}{2}\frac{\Delta \varepsilon }{\chi _{ms}^2 C(r_{ms} )^2}
\end{equation}

\noindent
where $\Delta \varepsilon $ is the additional radiative efficiency, $\chi
_{ms} = \sqrt {r_{ms} / M} $, and $C$ is general relativistic correction
factor defined by PT74.

Using this boundary condition we have the new function $f$ of radius
, that is

\begin{equation}
\label{eq4}
\begin{array}{l}
f = - \frac{d\Omega }{dr}(E^ + - \Omega L^ + )^{ - 2} \\ \\
\quad\quad [\int_{r_{ms} }^r {(E^+ } - \Omega L^ + )\frac{dL^ + }{dr}dr - \frac{3(E_{ms} - \Omega _{ms}
L_{ms}^ + )^2\Delta \varepsilon }{2\Omega _{ms} \chi _{ms}^2 C_{ms}^{1 / 2}
}]
\end{array} .
\end{equation}

\noindent
$E^{+ }$ and $L^{+}$ are the
specific energy and the specific angular momentum of a particle in the disc,
respectively, and they read

\begin{equation}
\label{eq5} E^\dag = {\left( {1 - 2\chi ^{ - 2} + a_ * \chi ^{ -
3}} \right)} \mathord{\left/ {\vphantom {{\left( {1 - 2\chi ^{ -
2} + a_ * \chi ^{ - 3}} \right)} {\left( {1 - 3\chi ^{ - 2} + 2a_
* \chi ^{ - 3}} \right)^{1 / 2}}}} \right.
\kern-\nulldelimiterspace} {\left( {1 - 3\chi ^{ - 2} + 2a_ * \chi
^{ - 3}} \right)^{1 / 2}},
\end{equation}

\begin{equation}
\label{eq6} L^\dag = M\chi {\left( {1 - 2a_ * \chi ^{ - 3} + a_ *
^2 \chi ^{ - 4}} \right)} \mathord{\left/ {\vphantom {{\left( {1 -
2a_ * \chi ^{ - 3} + a_ * ^2 \chi ^{ - 4}} \right)} {\left( {1 -
3\chi ^{ - 2} + 2a_ * \chi ^{ - 3}} \right)^{1 / 2}}}} \right.
\kern-\nulldelimiterspace} {\left( {1 - 3\chi ^{ - 2} + 2a_ * \chi
^{ - 3}} \right)^{1 / 2}}.
\end{equation}

\noindent
$\chi = \sqrt {r / M} $ is the dimensionless radial coordinate, and
$a_\ast = a / M$ is the dimensionless black hole spin parameter.

The power dissipated in the corona (i.e. the magnetic Poynting flux in the
vertical direction from the thin disc) is

\begin{equation}
\label{eq7}
Q_{cor} = P_m \upsilon _P = \frac{B^2}{8\pi }\upsilon _P
\end{equation}

\noindent
where $P_m $ is the magnetic pressure in the disc, and $\upsilon _P $ is the
velocity of magnetic flux transported vertically in the disc. Here we assume
the velocity $\upsilon _P $ of magnetic flux tubes is proportional to their
internal Alfven velocity, i.e. $\upsilon _P = b\upsilon _A $, and $b$ is
related to the efficiency of buoyant transport of magnetic structure in the
vertical direction inside the disc, which is of the order of unity for
extremely evacuated tubes \citep{b21}.

Now we give the equations of the disc structure as follows. The equation of
vertical pressure balance in the vertically-averaged form is \citep{b25}

\begin{equation}
\label{eq8}
H = (P / \rho )^{1 / 2}(r^3 / M)^{1 / 2}AB^{ - 1}C^{1 / 2}D^{ - 1 / 2}E^{ -
1 / 2}
\end{equation}

\noindent
where $H$ is the height of the accretion disc, $P$ and $\rho $ are pressure and
density of the disc respectively.  $A$,$B$, $C$,$D$,
$E$ are general relativistic correction factors defined as follows:

\begin{equation}
\label{eq9}
\left\{ {\begin{array}{l}
 A = 1 + a_\ast ^2 \chi ^{ - 4} + 2a_\ast ^2 \chi ^{ - 6} \\
 B = 1 + a_\ast \chi ^{ - 3} \\
 C = 1 - 3\chi ^{ - 2} + 2a_\ast ^2 \chi ^{ - 3} \\
 D = 1 - 2\chi ^{ - 2} + a_\ast ^2 \chi ^{ - 4} \\
 E = 1 + 4a_\ast ^2 \chi ^{ - 4} - 4a_\ast ^2 \chi ^{ - 6} + 3a_\ast ^4 \chi
^{ - 8} \\
 \end{array}} \right.
\end{equation}

The equation of energy conservation is (see Eq.(5.6.13) in NT73)

\begin{equation}
\label{eq10}
W = \frac{4}{3}(M / r^3)^{ - 1 / 2}CD^{ - 1}Q ,
\end{equation}

\noindent
and $W$ is integrated shear stress defined as

\begin{equation}
\label{eq11}
W = 2\int_0^h {t_{r\varphi } } dz \sim 2t_{r\varphi } H,
\end{equation}
\noindent where $t_{r\varphi } $ is the interior viscous stress in the disc.

The equation of state for gas on the disc is

\begin{equation}
\label{eq12}
P = P_{mag} + P_{tot} = P_{mag} + \frac{1}{3}aT^4 + \rho _0(T/m_p )
\end{equation}

\noindent
where $P_{tot} $ is the total pressure (gas pressure plus radiation
pressure) at disc mid-plane, $a$ is the radiative constant.  $ m_p $ is the rest mass of proton, $ \rho _0$ and $ T$ are the density of rest mass
and the temperature in the disc, respectively.

In the SSD model the interior viscous stress
$t_{r\varphi } $ is usually assumed to be proportional to the pressure, i.e.
\begin{equation}
\label{eq13}
t_{r\varphi} = \alpha P.
\end{equation}
\noindent

Since the process of generating magnetic fields in the accretion disc is
still unclear, we adopted different magnetic pressures as:

\begin{equation}
\label{eq14}
P_{mag} \simeq \left\{ {\begin{array}{l}
 \alpha_{0} P_{tot} = \alpha_{0} (P_{gas} + P_{rad} ) \\
 \alpha_{0} P_{gas} \\
 \alpha_{0} \sqrt {P_{gas} P_{tot} } \\
 \end{array}} \right.  ,
\end{equation}
 and as $ \alpha_{0} $ is a constant of the order of unity, we adopt $ \alpha_{0} = 1 $ in our calculations.
\noindent

In the disc-corona scenario, the soft photons from the disc are scattered by the electrons in the corona to X-ray bands. About half of
the scattered photons are intercepted by the disc. It is a rough approximation because light bending effects ought to be taken into account
at the vicinity of BH. The reflection albedo $a_0$ is relatively low $( a_0 \sim 0.1 - 0.2 )$, and most of the incident
photons from the corona are re-radiated as black body radiation \citep{b35}. Thus the
energy transport equation for the disc is

\begin{equation}
\label{eq15}
aT^4 = 2\kappa H\rho ( Q - Q_{cor} + \frac{1 - a_0 }{2}Q_{cor} ),
\end{equation}

\noindent
where $\kappa $ is the Rosseland mean of total opacity which can be expressed
as

\begin{equation}
\label{eq16}
\kappa = (0.64\times 10^{23})\rho T^{ - 7 / 2} + 0.40 \quad (cm^{2}/g),
\end{equation}

\noindent and $a_0 = 1.5$ is adopted in the calculations.

Solving equations (2),(7)-(16) numerically, the power dissipated in the
corona $Q_{cor} $ and the structure of the disc can be derived as function
of $r$. The ratio of the power dissipated in the corona to the total for such
disc-corona system is given as

\begin{equation}
\label{eq17}
\left\langle f \right\rangle = \frac{\int {Q_{cor} 2\pi \left( {{\varpi \rho }
\mathord{\left/ {\vphantom {{\varpi \rho } {\sqrt \Delta }}}
\right. \kern-\nulldelimiterspace} {\sqrt \Delta }}
\right)_{\theta = \pi \mathord{\left/ {\vphantom {\pi 2}} \right.
\kern-\nulldelimiterspace} 2} dr} }{\int {Q
2\pi \left( {{\varpi \rho }
\mathord{\left/ {\vphantom {{\varpi \rho } {\sqrt \Delta }}}
\right. \kern-\nulldelimiterspace} {\sqrt \Delta }}
\right)_{\theta = \pi \mathord{\left/ {\vphantom {\pi 2}} \right.
\kern-\nulldelimiterspace} 2} dr} } \quad ,
\end{equation}

\noindent
the concerned Kerr metric coefficients are given by \citep{b34}
\begin{equation}
\label{eq18} \left\{ {\begin{array}{l}
 \varpi = \left( {\Sigma \mathord{\left/ {\vphantom {\Sigma \rho }} \right.
\kern-\nulldelimiterspace} \rho } \right)\sin \theta ,\mbox{
}\Sigma ^2
\equiv \left( {r^2 + a^2} \right)^2 - a^2\Delta \sin ^2\theta , \\
 \rho ^2 \equiv r^2 + a^2\cos ^2\theta ,\mbox{ }\Delta \equiv r^2 + a^2 -
2Mr. \\
 \end{array}} \right.
\end{equation}

\section{ NUMERICAL RESULTS}

\begin{figure}
\vspace{0.5cm}
\begin{center}
\includegraphics[width=6cm]{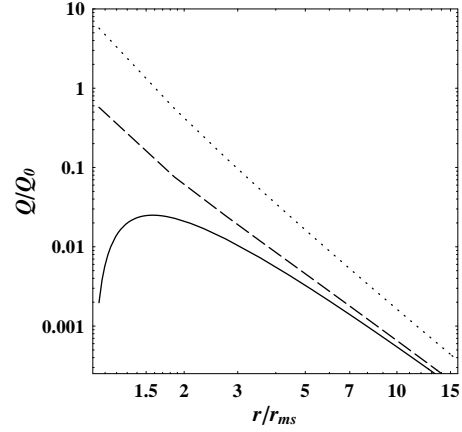}
 \caption{The value of $Q / Q_0 $ varying with  $r / r_{ms} $ for the different values of $\Delta \varepsilon
$ : $\Delta \varepsilon = 0$ (solid lines), $\Delta
\varepsilon = 0.1$ (dashed lines), $\Delta \varepsilon = 1$ (dotted lines); The dimensionless BH spin parameter  $a_\ast = 0.95$ is assumed.}
\end{center}
\end{figure}

\noindent
By using Eq.(2)-(6) we have the total gravitational power $ Q $ versus the radius of the disc for the different values of  $\Delta \varepsilon
$ as shown in Fig.1. Note that the power $ Q $ is in units of $Q_0 = \dot {M} / M^2 $ , where $M $ is the BH mass. Inspecting the above Fig, we have the following results: (I) the power $ Q $ depends sensitively on the values of $\Delta \varepsilon $, i.e. the
magnetic torque can considerably enhance the amount of energy $ Q $ released in the disc; (II) the power $ Q $ decreases monotonically with the
increasing radius, it indicates that the energy flux mostly comes from the inner region of the disc.

The ratio of accretion power dissipated into the corona $ \left\langle f \right\rangle
$ as functions of the parameter $\Delta \varepsilon $ and the dimensionless BH spin parameter $a_\ast $ are shown in Fig.2 and Fig.3(a), and the magnetic pressure $P_{mag} = \alpha_{0} P_{tot} $ is adopted in our calculations.
From the above Fig.2 and Fig.3(a), we find that the global value of $ \left\langle f \right\rangle $ is insensitive on the parameter $\Delta \varepsilon $ for $\Delta \varepsilon > 1 $, but the global value increases with the increasing the spin parameter $a_\ast $.
  The curves of $ \left\langle f \right\rangle $ versus accretion rate $\dot{m} $ for the deferent BH spin parameter $a_\ast $ are plotted in Fig3.(b).
The dimensionless accretion rate $\dot{m} $ is defined as $ \dot{m} = \dot{M}/\dot{M}_{Edd}$ in the calculations.  As shown in Fig3b the global value $ \left\langle f \right\rangle $ is independent of accretion rate $\dot{m} $ for a given BH spin parameter $a_\ast $ if the magnetic pressure $P_{mag} = \alpha_{0} P_{tot} $ is adopted. We find that this result is consistent with those in \citet{b5}, though the central black hole is Kerr black
hole in our model, while a Schwarzschild black hole in \citet{b5}.


\begin{figure}
\begin{center}
\includegraphics[width=6cm]{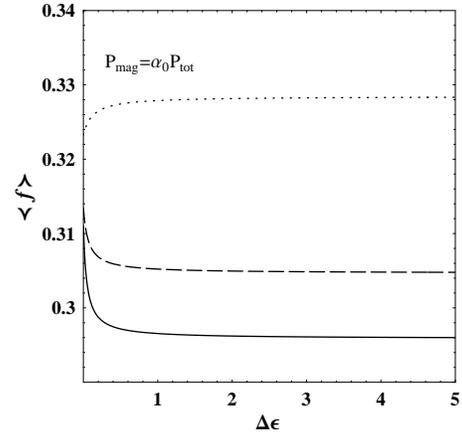}
\caption{The global value of $ \left\langle f \right\rangle $ as a function of the
 parameter $\Delta \varepsilon $ for three different values
of $a_\ast $: $a_\ast = 0$ (solid lines), $a_\ast = 0.5$ (dashed lines), $a_\ast = 0.9$ (dotted lines). $ \dot{m} = 0.1$,  $ \alpha = 0.3 $ is adopted in the calculations.}
\label{fig2}
 \end{center}
\end{figure}

\begin{figure}
\vspace{0.5cm}
\begin{center}
{\includegraphics[width=6cm]{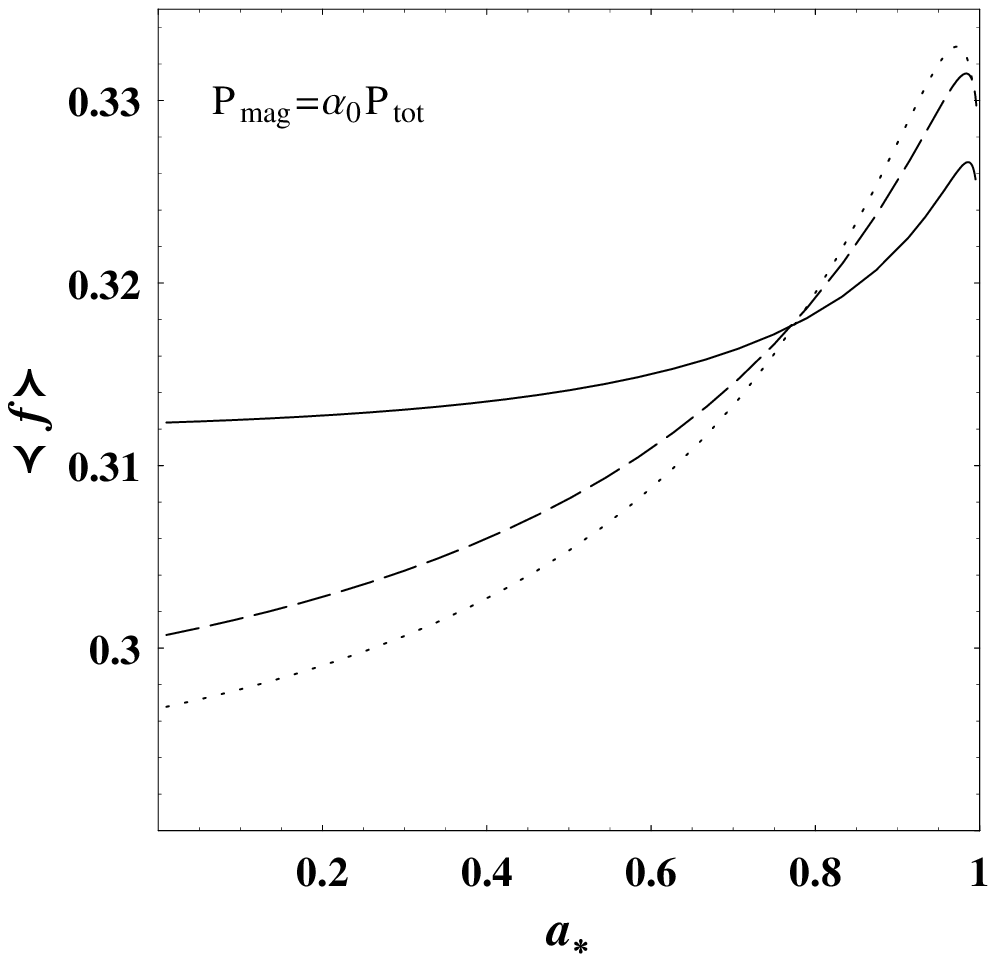}
 \centerline{(a)}
 \includegraphics[width=6cm]{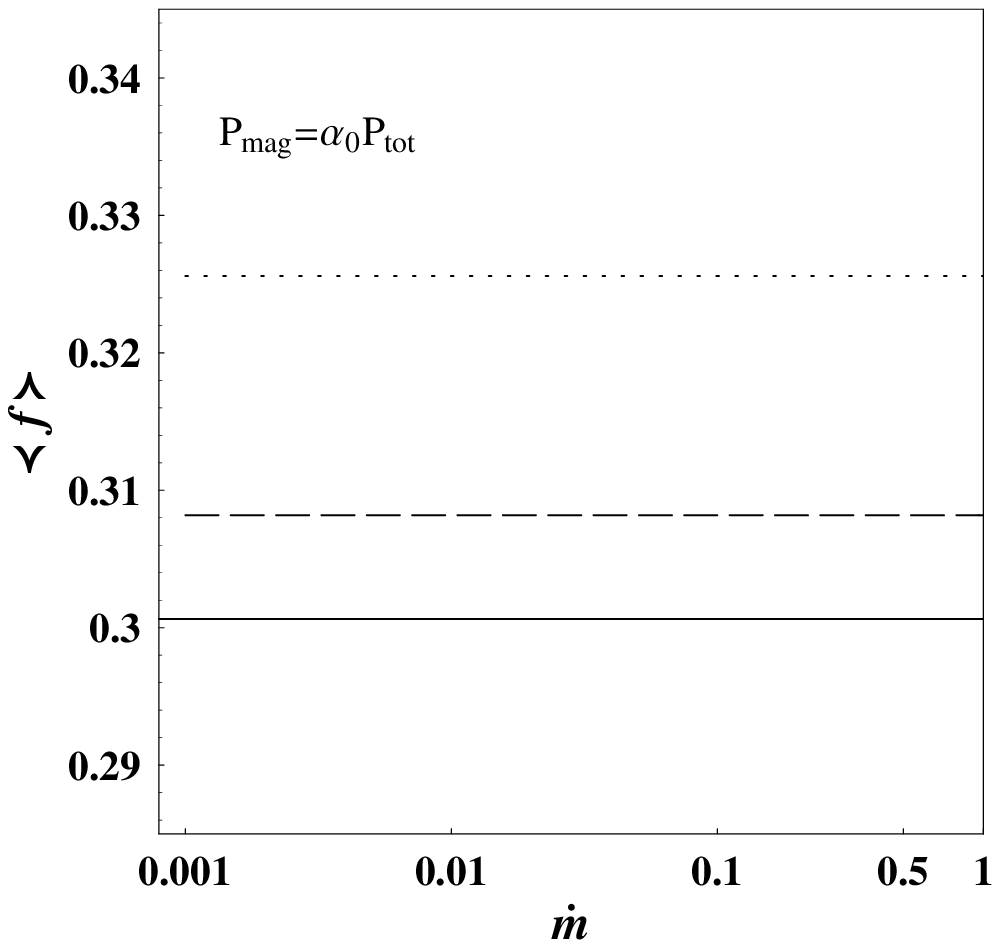}
 \centerline{(b)}}
 \caption{(a) The global value of $ \left\langle f \right\rangle $ varies with $a_\ast
$ for different values of $\Delta \varepsilon $: $\Delta \varepsilon =
0$ (solid lines), $\Delta \varepsilon = 0.1$ (dashed lines), $\Delta
\varepsilon = 1$ (dotted lines). $\dot{m} = 0.1$ is adopted in the calculations, (b) The global value of $
\left\langle f \right\rangle $ varies with $\dot{m} $ for different values of $a_\ast $: $a_\ast = 0$(solid lines), $a_\ast = 0.5$ (dashed lines), $a_\ast = 0.9$ (dotted lines), $\Delta \varepsilon = 0.1$
and $ \alpha = 0.3 $ are adopted in the calculations.}\label{fig3}
\end{center}
\end{figure}

We plot the curves of the global value $ \left\langle f \right\rangle $ versus different parameters
with $P_{mag} = \alpha_{0} P_{gas} $ and $P_{mag} = \alpha_{0} \sqrt {P_{gas} P_{tot}
} $ in Fig 4 and Fig 5, respectively.

Inspecting the Fig.4 and Fig.5 we find that our disc-corona model calculations with $P_{mag} = \alpha_{0} P_{gas}$ show the integration fraction of accretion power dissipated into the corona $ \left\langle f \right\rangle \sim 0.270 - 0.298$ with the BH spin parameter $a_\ast \sim 0.01 - 0.998$, and the model with
$P_{mag} = \alpha_{0} \sqrt {P_{gas} P_{tot} } $ shows that $ \left\langle f \right\rangle \sim 0.285 -0.315$ with the BH spin parameter $a_\ast \sim 0.01 - 0.998$
at $\dot{m} < 0.05$, for different values of $\Delta\varepsilon$.

 From Fig.4(a) we find that the values of $ \left\langle f \right\rangle$ depend on the parameters $a_\ast$ and $\Delta\varepsilon$ when $\Delta\varepsilon < 1$. $\left\langle f \right\rangle$ decreases (increases) as $ \Delta\varepsilon $ changes from 0 to 1, for $ a_{\ast} < 0.4 $ ( $ a_{\ast} > 0.4 $ ). In fact, the magnetic fields connecting the plunging region to the disc can exert stresses on the inner edge of the disc and transfer energy from the plunging region to the disc-coronae system. So the values of $ Q_{cor}$ and $ Q $ all increase with the increasing $ \Delta\varepsilon$. $ Q_{cor}$ and $ Q $ are also positively correlated to BH spins. For $ a_{\ast} < 0.4 $, the value of $ Q_{cor}$ increases more slowly than that of $ Q $ with the increasing $ \Delta\varepsilon$. Thus the ratio of accretion power dissipated into the corona, $ \left\langle f \right\rangle$, decreases with the increasing $ \Delta\varepsilon $. For $ a_{\ast} = 0.4 $, we find that the change of the magnetic torque has no influence on the ratio of the power dissipated in the corona to the total power. This critical BH spin parameter is about $ a_{\ast} = 0.78 $ in Fig.3(a), and is about $ a_{\ast} = 0.6 $ in Fig.5(a).

 In addition, it should be noted that the value of $ \left\langle f \right\rangle$  is also insensitive on the $\Delta \varepsilon$ for $\Delta\varepsilon > 1$, as $P_{mag} = \alpha_{0} P_{gas}$ or $P_{mag} = \alpha_{0} \sqrt {P_{gas} P_{tot}
} $ is adopted in our calculations.

 As shown in Fig.4(b) and Fig.5(b), the value $ \left\langle f \right\rangle$ decreases monotonically with the increasing accretion rate for $\dot{m} \geq 0.05$ . $ \left\langle f \right\rangle$ can reach $ \sim 0.3$ at the lower accretion rates. So it seems that the low luminosity BH can be fitted by our disc-corona model nicely. \citet{b21} also proposed that the magnetic corona should be stronger at low accretion rates, and their strength depends upon the nature of magnetic dissipation inside the disc.


\begin{figure}
\vspace{0.5cm}
\begin{center}
{\includegraphics[width=6cm]{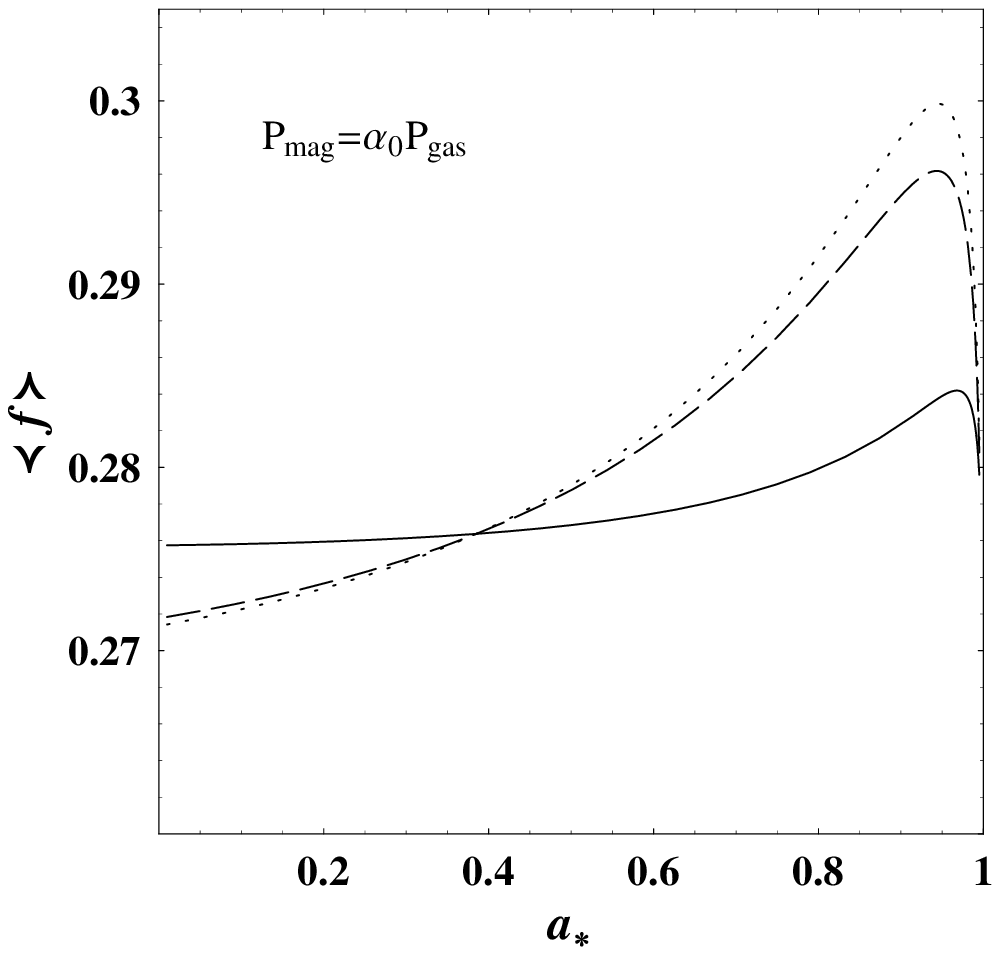}
 \centerline{(a)}
 \includegraphics[width=6cm]{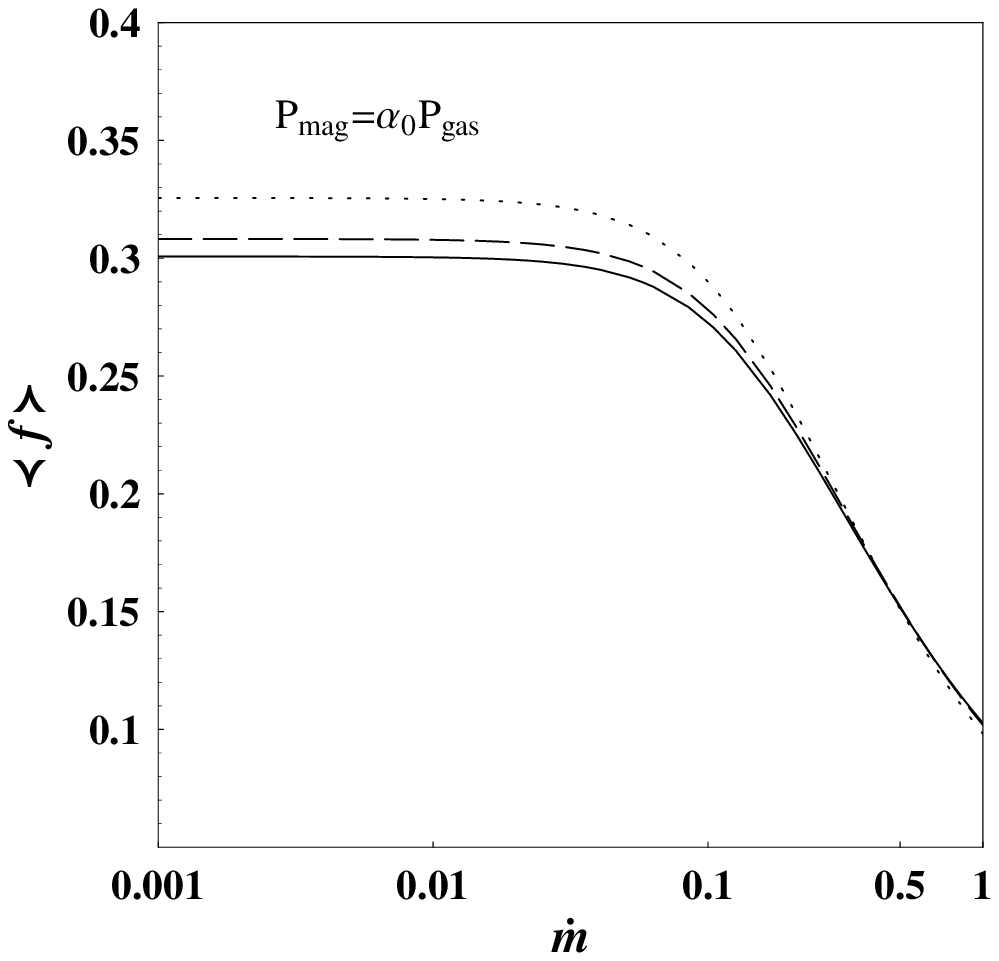}
 \centerline{(b)}}
 \caption{(a) The global value of $ \left\langle f \right\rangle $ varies with $a_\ast
$ for different values of $\Delta \varepsilon $: $\Delta \varepsilon =
0$ (solid lines), $\Delta \varepsilon = 0.1$ (dashed lines), $\Delta
\varepsilon = 1$ (dotted lines). $\dot{m} = 0.1$ is adopted in the calculations, (b) The global value of $
\left\langle f \right\rangle $ varies with $\dot{m} $ for different values of $a_\ast $: $a_\ast = 0$(solid lines), $a_\ast = 0.5$ (dashed lines), $a_\ast = 0.9$ (dotted lines), $\Delta \varepsilon = 0.1$
and $ \alpha = 0.3 $ are adopted in the calculations.}\label{fig4}
\end{center}
\end{figure}


\begin{figure}
\vspace{0.5cm}
\begin{center}
{\includegraphics[width=6cm]{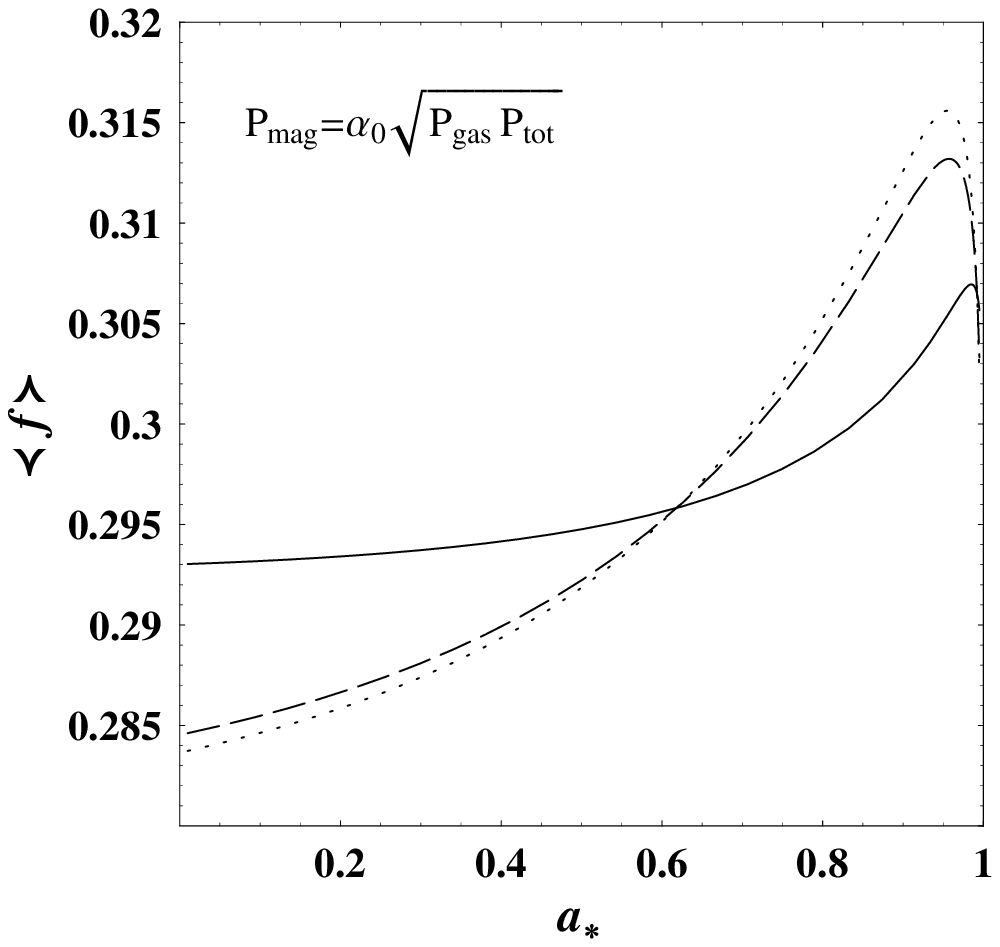}
 \centerline{(a)}
 \includegraphics[width=6cm]{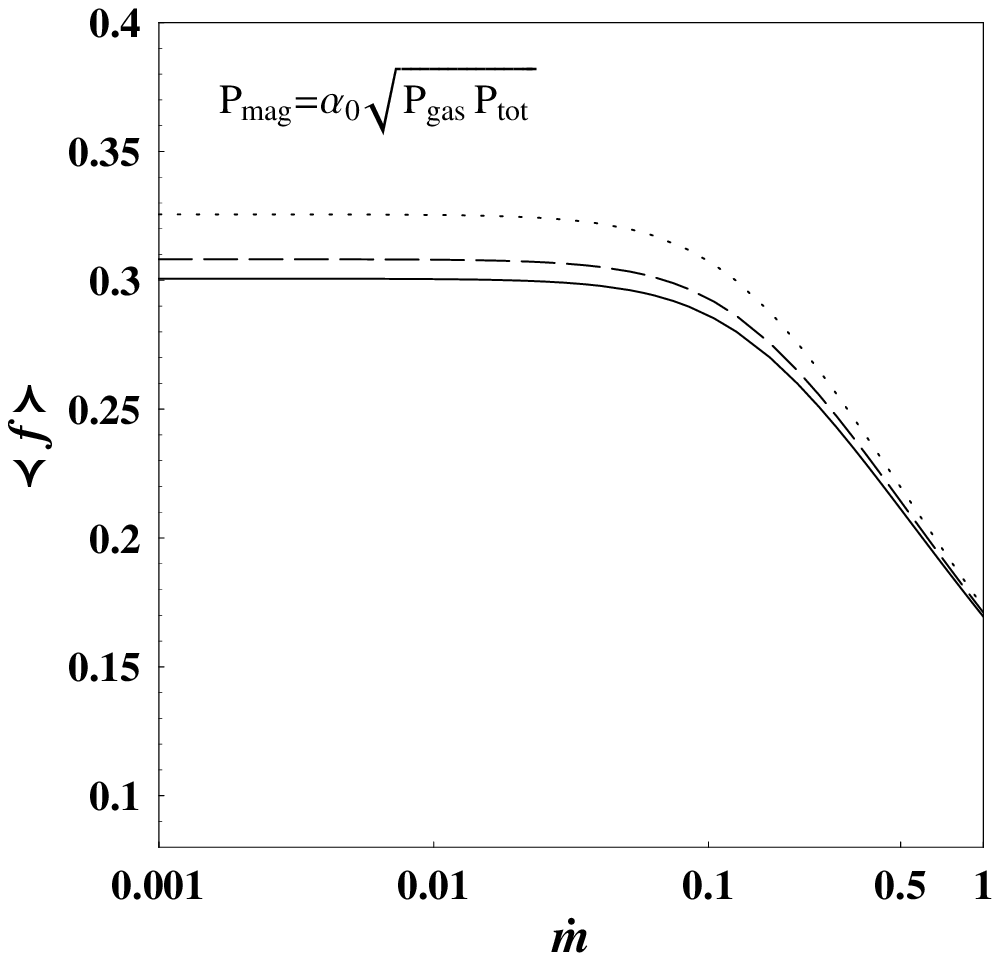}
 \centerline{(b)}}
 \caption{(a) The global value of $ \left\langle f \right\rangle $varies with $a_\ast
$ for different values of $\Delta \varepsilon $: $\Delta \varepsilon =
0$ (solid lines), $\Delta \varepsilon = 0.1$ (dashed lines), $\Delta
\varepsilon = 1$ (dotted lines). $\dot{m} = 0.1$ is adopted in the calculations, (b) The global value of $
\left\langle f \right\rangle $ varies with $\dot{m} $ for different values of $a_\ast $ : $a_\ast = 0$ (solid lines), $a_\ast = 0.5$ (dashed lines), $a_\ast = 0.9$ (dotted lines), $\Delta \varepsilon = 0.1$ and $ \alpha = 0.3 $ are
adopted in the calculations.}\label{fig5}
\end{center}
\end{figure}

\section{SIMULATION OF DISC SPECTRUM}

\noindent
In the disc-corona scenario, the comptonized spectrum has been computed by some authors using different
approaches in previous works. Two kinds of approaches
are used in order to get the emergent spectrum. One of common approaches is to solve the radiative transfer equation either numerically or analytically \citep{b32,b28}. Another kind of approach is the Monte Carlo simulation\citep{b27,b7,b31,b8,b18} . Recently, \citet{b18} got the output spectra in the
cases with and without the magnetic coupling effects ( Li 2000, Li \& Paczy\'{n}shi 2000, Li 2002a, Li 2002b ) by using the Monte
Carlo simulation.

In this paper, the steps of our simulations are: (i) sample a seed photon
from the cold disc; (ii) draw a value for its free path and test whether it
can leave the disc or corona; (iii) simulate the interaction of the photon
with the medium; (iv) repeat steps (ii), (iii) till the photon leaves the
system of the corona and disc.

In our disc-corona scenario, the dissipated power in the unit area of the
disc, $F$ is related to $Q $ by $F = Q - Q_{cor} $, and according to
Stefan-Boltzmann law we have the local effective temperature on the disc
expressed by

\begin{equation}
\label{eq19}
T_d (r) = (F / \sigma _{SB} )^{1 / 4},
\end{equation}

\noindent
where $\sigma _{SB} $ is the Stefan-Boltzmann constant. The local radiation
spectrum is defined by the Planck function:

\begin{equation}
\label{eq20}
B_\nu (r) = \frac{2h}{c^2}\frac{\nu ^3}{\exp [h\nu / k T_d (r)] - 1},
\end{equation}

\noindent
thus the multicolor black-body spectrum of the disc is

\begin{equation}
\label{eq21}
L_\nu = \int_{r_{in}} ^{r_{out}} {B_\nu } (r)2\pi rdr.
\end{equation}

The probability density of the seed photons can be written as

\begin{equation}
\label{eq22}
P(r,E ) = \tilde {P}(r)P(E),
\end{equation}

\noindent
where $\tilde {P}(r) = L(r)dr/L $ is the probability of a photon
emitted in the ring $r \sim r + dr$, and $L$ is the total luminosity of the disc. $P(E)$ is the number density of photons having an energy $ E = h \nu$ that is given by \citet{b27}

\begin{equation}
\label{eq23}
P( E ) = \frac{1}{2\zeta(3)}b^{3}E^{2}(e^{bE}-1)^{-1},
\end{equation}

\noindent
where $ b = 1 / T_d $ , $ \zeta(3) = 1.202 $ is the Riemann Zeta function.

 In our simulations,if the seed photon is scattered by the electron in the corona, the optical depth that the photon travels between the i-th and (i+1)-th scatterings can be drawn with $ \tau_{i} = - \ln \lambda$, where $ 0 \leq \lambda \leq 1 $ is a random number corresponding to a random event. The free path of photon can be drawn with

\begin{equation}
\label{eq24}
\frac{\tau_{i}}{n_{e} \sigma} = \frac{-\ln\lambda \sigma_{T}H_{c}}{\sigma \tau_{c}},
\end{equation}
 where $ n_{e}$ , $\sigma $ and $ \sigma_{T} = 6.65 \times 10^{-24} cm^{2}$ are the number density of the electrons, cross-sections of
 scattering and Thomson scattering, respectively. $ H_{c}$ and $ \tau_{c}$ are the vertical height and optical depth of the corona. Since
 the electrons in the hot corona are relativistic, the cross-section of scattering $\sigma$ depends not only on the energy of the photon
 but also on the energy and direction of the electron. So we use the cross-section averaged over the distribution of the electrons in \citet{b8}to draw the free path of the photon.

 In the disc-corona system, hard X-ray photons irradiating the disc from the corona can be absorbed by the atoms in the disc, as well as being scattered by the free electrons. In this case, we choose the lesser of the two free paths that are drawn with $ \sigma_{a}$ and $\sigma_{s}$ to draw the free path of the photon, where $ \sigma_{a}$ and $\sigma_{s}$ are respectively the cross sections of absorption and scattering. The cross sections of absorption $\sigma_{a}$ is taken from \citet{b22}, and the cross sections of
 scattering can be given by the Klein-Nishina formula

 \begin{equation}
\label{eq25}
\ \sigma_{s} = \frac{3\sigma_{T}}{4\varepsilon}[(1 - \frac{4}{\varepsilon} - \frac{8}{\varepsilon^{2}}) \ln(1 + \varepsilon) + \frac{1}{2} + \frac{8}{\varepsilon} - \frac{1}{2(1 + \varepsilon)^{2}}],
\end{equation}

\noindent where $ \varepsilon $ is the energy of the incident photon in unit of electron-rest-energy.

 As the free path is known, the position that the photon arrives before the next interaction can be calculated. If the photon is outside the disc-corona system, it will escape away from the system and its energy and direction are recorded. If the photon transfers from the disc to the corona or inverse, the point where the trajectory of the photon cross the interface between the disc and corona will be regarded as the next initial position of the photon in calculations.

 In the simulating the interaction of photon and electron, we can sample a electron from thermal distribution, then calculate the energy and direction of scattered photon. In the calculation, we follow the calculation procedure described in P83.The bound-free absorption of hard X-rays by the atoms in the disk will lead to ionization and vacancy, and induce emission of fluorescence lines with the probability called fluorescence yield, $ Y $. In our simulations, if the photon is absorbed by a certain atom or ion, we can draw a random number $ \lambda $ ( $ 0 \leq \lambda \leq 1 $) and compare it with the corresponding fluorescence yield. If $ \lambda \leq Y $, an emission line is brought out, whose direction can be sampled from the isotropic distribution. If $ \lambda > Y$, the photon vanishes and its trajectory ends. We only consider the 6.4 keV Fe $K_{\alpha} $ fluorescence line in our simulations, due to the large abundance and cross-section of absorption of iron. If the energy of the absorbed photon is less than the iron K-shell absorption edge, i.e. E $ < $ 7.1 keV, no emission line is produced, while Fe $K_{\alpha}$ lines emanate with $ Y = 0.34 $ for E $> $ 7.1 keV \citep{b18}.

 In our stimulations, the half height of the disc is assumed to be $H = 0.05r_{ms}$. All the spectra in this paper are given in forms of curves rather than histogram to make the curves look smooth. Fortran 95 is used in the calculations and Mathematica 7.0 software is adopted to plot the figures.

The typical results for the spectra of our Monte Carlo simulations are given
in Fig. 6. The spectrum of the disc-corona system comprises three
components, the black-body spectrum formed by unscattered photons, the
power-law spectrum formed by photons that escape from the corona after
several times of inverse Compton scatterings, and the reflected spectrum
characterized by the iron fluorescence line and reflection hump.


\begin{figure}
\begin{center}
\includegraphics[width=6cm]{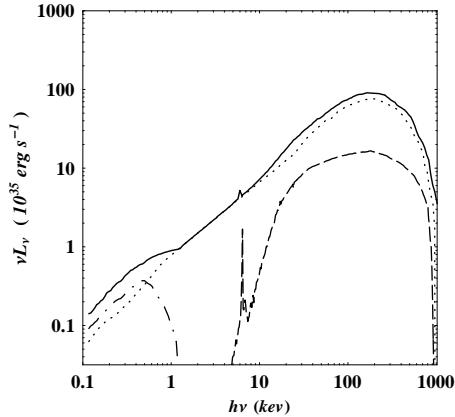}
\caption{The emerged spectrum from the disc-corona system. The total
emissive spectrum and its black body,power-law and reflected components are shown in solid , dot-dashed, dotted
and dashed lines, respectively. The radii of the inner and outer edges of the corona  are taken as $r_{in} = r_{ms} $ and $r_{out} = 100 r_{ms}$,  and $a_\ast = 0.998$, $\Delta \varepsilon = 0.1$ and $ \dot{m} = 0.05 $ are
adopted in the calculations.}\label{fig6}
 \end{center}
\end{figure}

\begin{figure}
\vspace{0.5cm}
\begin{center}
{\includegraphics[width=6cm]{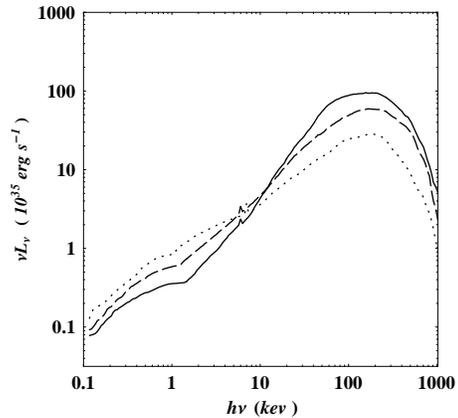}
 \centerline{(a)}
 \includegraphics[width=6cm]{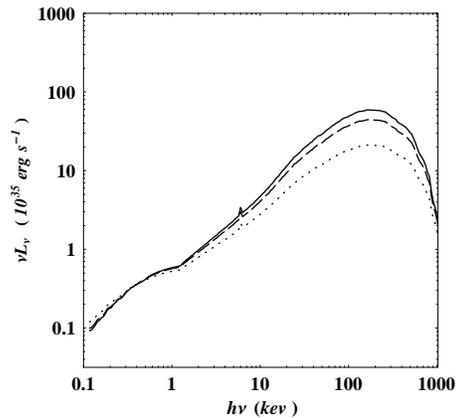}
 \centerline{(b)}}
\caption{The simulated spectra of the disc-corona system.
(a) Solid, dashed, and dotted lines correspond to the spectra with different
heights of corona : $H_C = 2 r_{ms}$, $H_C = r_{ms}$, $H_C = 0.5 r_{ms}$,
for the radii of the inner and outer edges of the corona  are taken as $r_{in} = r_{ms} $ and $r_{out} = 100 r_{ms}$ . (b) Solid, dashed, and dotted lines correspond to the spectra with different
the radii of the outer edges of the corona  : $r_{out} = 100 r_{ms}$, $r_{out}= 50 r_{ms}$, $ r_{out} = 20 r_{ms}$,
for $r_{in} = r_{ms} $ and $H_C = r_{ms}$.  $a_\ast = 0.5$, $\Delta \varepsilon = 0.1$ and $ \dot{m} = 0.05 $ are
adopted in the calculations.}\label{fig7}
 \end{center}
\end{figure}

\begin{figure}
\vspace{0.5cm}
\begin{center}
{\includegraphics[width=6cm]{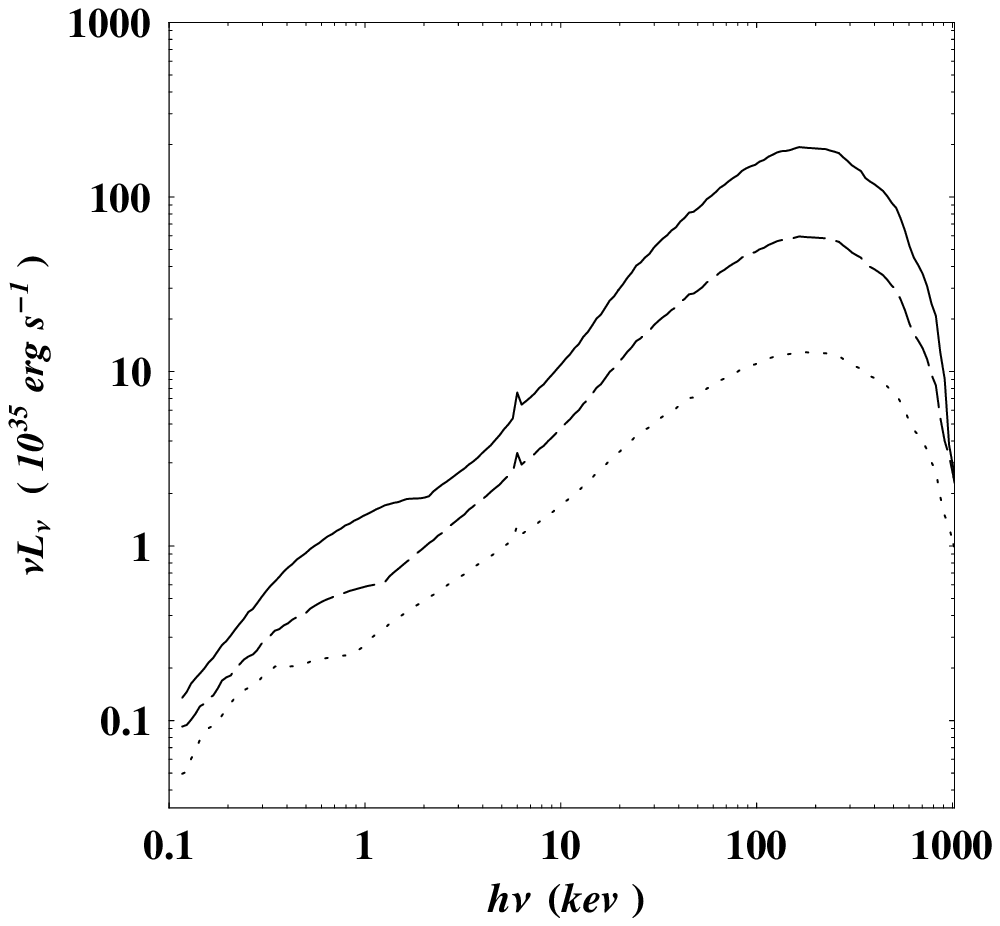}
 \centerline{(a)}
 \includegraphics[width=6cm]{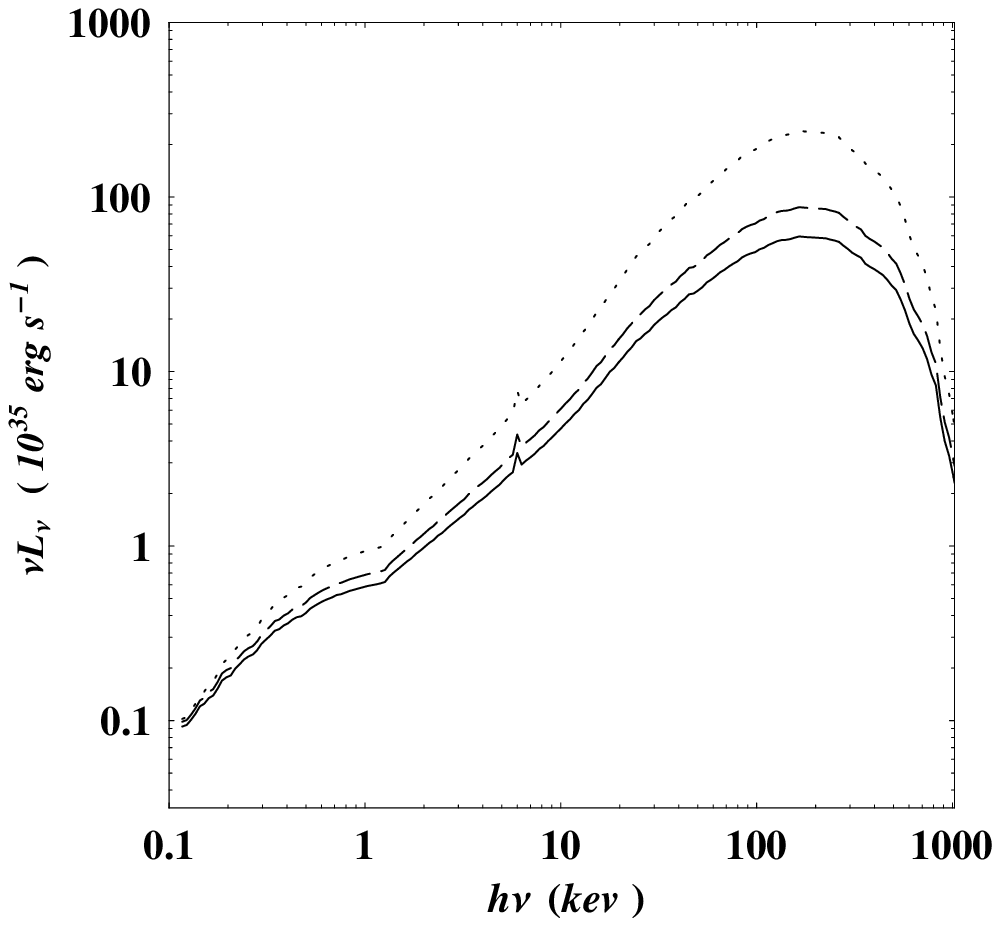}
 \centerline{(b)}
 \includegraphics[width=6cm]{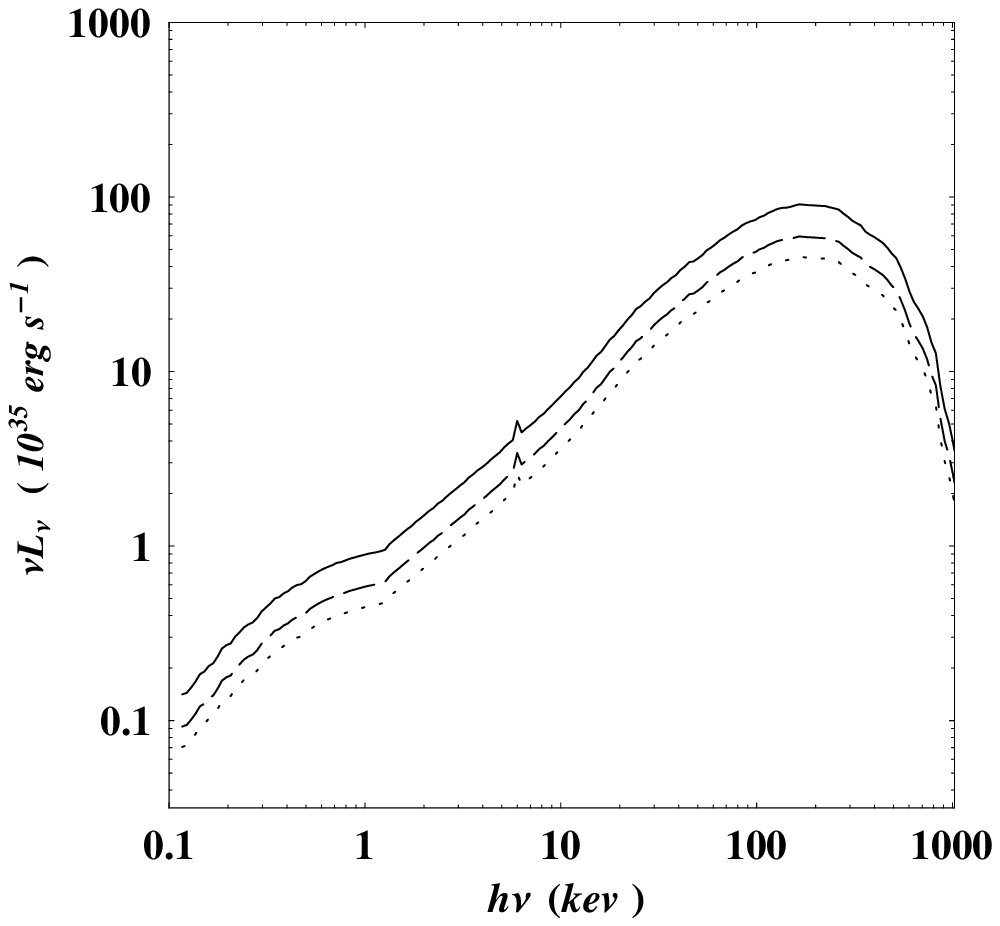}
 \centerline{(c)}}
 \caption{The simulated spectra of the disc-corona system.
(a) Solid, dashed, and dotted lines correspond to the spectra with different
accretion rates  : $\dot{m} = 0.1 $, $\dot{m} = 0.05 $, $\dot{m} = 0.01$,
for $ a_\ast = 0.5$, $\Delta \varepsilon = 0.1$; (b) Solid, dashed, and dotted lines correspond to the spectra with different
 $\Delta \varepsilon$ : $\Delta \varepsilon = 0.1$, $\Delta \varepsilon = 1$, $\Delta \varepsilon = 2 $, for $ a_\ast = 0.5$, $\dot{m} = 0.05 $ ; (c) Solid, dashed, and dotted lines correspond to the spectra with different
spin parameters :  $a_\ast = 0.998$, $a_\ast = 0.5$, $a_\ast = 0.1$, for $\dot{m} = 0.05 $, $\Delta \varepsilon = 0.1$ .  $H_C = r_{ms}$, $r_{in} = r_{ms} $ and $r_{out} = 100 r_{ms}$ are
adopted in the calculations.}\label{fig8}
\end{center}
\end{figure}

Spectra with different coronal geometry parameters are given in Fig. 7. We stimulate spectra with different heights of corona: $H_C = 0.5 r_{ms}$, $H_C = r_{ms}$, $H_C = 2 r_{ms}$, in Fig. 7(a). It is found that the spectra steepen in the 2-200 keV range with the increasing height of corona. This result can be explained as follows: when the corona  becomes thicker, the amount of photon-electron scattering increases, and the energy of the escaped photons increases. Therefore, hardening of the spectrum is anticipated.
We present the spectra with different radii of the outer edge of the corona : $r_{out} = 100 r_{ms}$, $r_{out}= 50 r_{ms}$, $ r_{out} = 20 r_{ms}$  in Fig.7(b). The fluxes increase with $r_{out}$, since a larger disc can lead to more energy from the coronal surface. The spectral profile changes slightly in this case.

Spectra with different accretion rates $\dot{m}$, $ \Delta \varepsilon $ and BH spin parameters $a_\ast $ are given in Fig. 8. From Fig. 8(a) we find that the fluxes increase as $\dot{m}$ changes from 0.01 to 0.1, and the spectral profile changes obviously. As the accretion rate $\dot{m}$ increases, the total gravitational energy dissipated is enhanced remarkably. In this case, more seed photons escape from the surface of the disc; these photons also have higher energies. It is found in Fig.8(b) that the fluxes increase as $\Delta \varepsilon $ increases from 0.1 to 2, since the magnetic torque can transfer energy from the plunging region to the disc-corona system. The energy distribution of the seed photons, hence the emerged spectrum, also changes in this case. From Fig.8(c) we see that the fluxes increase with $ a_\ast $. Indeed, as $a_\ast $ increases the inner edge of disc approaches the BH horizon $ r_H = M( 1+ \sqrt{1 - a^{2}_{*}} )$ , and more gravitational energy can be released. $ M = 10 M_{\odot} $, and $P_{mag} = \alpha_{0} \sqrt {P_{gas} P_{tot}} $ is adopted through the calculations.

\section{DISSCUSSION}

\noindent
In this work, we propose a disc-corona model, in which magnetic fields
exert a torque on the inner edge of the accretion disc, and part of
gravitational energy is dissipated in the hot corona. The total
gravitational power $ Q $ is derived from the thin-disc dynamics equations and the global solutions are obtained
numerically. It is found that the fraction of the power dissipated into the corona in the total for such disc-corona system increases with the increasing the dimensionless black hole spin parameter $a_\ast $, but is
insensitive on the parameter $\Delta \varepsilon $ for $\Delta\varepsilon > 1$.

We simulate the emerged spectra from the disc-corona system for the different parameters using the Monte-Carlo method.
It is found that the spectral profile changes obviously with varying the height of the corona and the accretion rate of the disc.
The reasonable geometry of the corona is important for simulating the emerged spectra using the Monte-Carlo method. We adopt a slab corona in this paper. However the geometry of the corona is still matter of debate, e.g. \citet{b31} proposed a patchy corona that made of a number of separate active regions. Furthermore, the observed correlation between the photon index and the reflection strength \citet{b35} has demonstrated the need for further geometrical/dynamical parameters, such as the relativistic bulk motion velocity of the coronal material \citep{b19}.

Our model needs to be improved in other aspects. For example, the gravitational effects on the trajectories of photons need to be taken into account, the cooling of synchrotron radiation should be considered and the ray-tracing should be used in our calculations.

The magnetic pressure $P_{mag} $ plays an important role in our model. In
fact we can get the  magnetic energy density from the coronal power $Q_{cor}$ as the coronal geometry is assumed. Then the small-scale magnetic field $ Q_{dynamo}$ in the corona can be derived from the  magnetic energy density.

On the other hand, models and simulations of jet production \citep{b3,b4,b20}show that it is the
poloidal component of the large-scale magnetic field which mainly drives the production
of powerful jets. Several theoretical models have been proposed for acceleration and
collimation of jets, which can be divided into two main regimes, the
Poynting flux regime and the hydromagnetic regime. Both regimes are related
to a poloidal magnetic field threading the disc or BH, from which energy and
angular momentum are extracted. In the Poynting flux regime, energy is
extracted in Poynting flux (i.e. purely electromagnetic energy), but in the
form of magnetically driven material winds in the latter regime. Furthermore, some authors have agreed that jet formation should involve an
accretion disc threaded by a large-scale magnetic field \citep{b17,b20}.

Though the origin of large-scale magnetic field is still under controversy,
some previous works \citep{b33,b29} have
proposed that the large-scale field can be produced from the small-scale
field created by dynamo processes. The length scale of the fields created by
dynamo processes is of the order of the disc thickness $H$, and the poloidal
component of the magnetic field is given approximately by
\begin{equation}
\label{eq26}
B_P \sim (H /
r)B_{dynamo},
\end{equation}
\noindent
If the field is created in the thin accretion discs ($H \ll
r)$, the large-scale field is very weak. For the ADAF cases, the disc
thickness $H \sim r$ and the poloidal component of the magnetic field shall
be stronger. In our disc-corona scenario, the energetically dominant corona
are the ideal sites for launching the powerful jets/outflows \citep{b21}. The large-scale magnetic fields created by dynamo processes in
the corona are significantly stronger than the thin disc due to the corona being much thicker than the cold, thin disc. So the
corona can power a stronger jet than the thin disc. We shall discuss the
acceleration of jet in the Poynting flux regime and the hydromagnetic regime
in future work.

\section*{Acknowledgments}

This work is supported by the National Basic Science Foundation of China under the Distinguished Young Scholar Grant 10825313, the National Basic Research Program of China (2009CB4901) and the Project for Excellent Young and Middle-Aged Talent of Education Bureau of Hubei Province under Grant Q200712001. We are very grateful to the anonymous referee for his/her helpful comments on the manuscript.

\clearpage

\end{document}